\documentclass[preprint,12pt]{elsarticle}
\usepackage{lineno}
\modulolinenumbers[5]
\usepackage{amsmath,amssymb,mathrsfs}
\usepackage{graphicx}
\usepackage{algorithm}
\usepackage{algorithmic}
\usepackage{tabularx}
\usepackage{cancel}
\usepackage{multirow}
\usepackage{color}
\usepackage{adjustbox}

\journal{Computer \& Geosciences}
\begin{document}
\begin{frontmatter}
	
\title{Performance prediction of finite-difference solvers for different computer architectures}


\author[1]{Mathias Louboutin \corref{mycorrespondingauthor}}
\ead{mloubout@eos.ubc.ca}
\author[2]{Michael Lange}
\author[1]{Felix J. Herrmann}
\author[2]{Navjot Kukreja}
\author[2]{Gerard Gorman}
\address[1]{Seismic Laboratory for Imaging and Modeling (SLIM),The University of BritishColumbia}
\address[2]{Earth Science and Engineering department,Imperial college, London}



%




\begin{keyword}
Finite-differences \sep HPC \sep Modelling \sep Multi-physics \sep Performance \sep Wave-equation
\end{keyword}

\begin{abstract}

The life-cycle of a partial differential equation (PDE) solver is often
characterized by three development phases: the development of a stable
numerical discretization; development of a correct (verified) implementation;
and the optimization of the implementation for different computer
architectures. Often it is only after significant time and effort has been
invested that the performance bottlenecks of a PDE solver are fully understood,
and the precise details varies between different computer architectures. One
way to mitigate this issue is to establish a reliable performance model that
allows a numerical analyst to make reliable predictions of how well a numerical
method would perform on a given computer architecture, before embarking upon
potentially long and expensive implementation and optimization phases. The
availability of a reliable performance model also saves developer effort as it
both informs the developer on what kind of optimisations are beneficial, and
when the maximum expected performance has been reached and optimisation work
should stop. We show how discretization of a wave-equation can be theoretically
studied to understand the performance limitations of the method on modern
computer architectures. We focus on the roofline model, now broadly used in the
high-performance computing community, which considers the achievable
performance in terms of the peak memory bandwidth and peak floating point
performance of a computer with respect to algorithmic choices. A first
principles analysis of operational intensity for key time-stepping
finite-difference algorithms is presented. With this information available at
the time of algorithm design, the expected performance on target computer
systems can be used as a driver for algorithm design.

\end{abstract}

\end{frontmatter}

\linenumbers
\section{Introduction}

The increasing complexity of modern computer architectures means that
developers are having to work much harder at implementing and optimising
scientific modelling codes for the software performance to keep pace with the
increase in performance of the hardware. This trend is driving a further
specialisation in skills such that the geophysicist, numerical analyst and
software developer are increasingly unlikely to be the same person. One problem
this creates is that the numerical analyst makes algorithmic choices at the
mathematical level that define the scope of possible software implementations
and optimizations available to the software developer. Additionally, even for
an expert software developer it can be difficult to know what are the right
kind of optimisations that should be considered, or even when an implementation is
"good enough" and optimisation work should stop. It is common that performance
results are presented relative to a previously existing implementation, but
such a relative measure of performance is wholly inadequate as the reference
implementation might well be truly terrible. One way to mitigate this issue is
to establish a reliable performance model that allows a numerical analyst to
make reliable predictions of how well a numerical method would perform on a
given computer architecture, before embarking upon potentially long and
expensive implementation and optimization phases. The availability of a
reliable performance model also saves developer effort as it both informs the
developer on what kind of optimisations are beneficial, and when the maximum
expected performance has been reached and optimisation work should stop.

Performance models such as the roofline model by~\citep{williams2009roofline}
help establish statistics for best case performance --- to evaluate the
performance of a code in terms of hardware utilization (e.g. percentage of peak floating point performance) instead of a relative speed-up. Performance models that establish algorithmic optimality and provide
a measure of hardware utilization are increasingly used to determine effective
algorithmic changes that reliably increase performance across a wide variety of
algorithms~\citep{asanovic2006landscape}. However, for many scientific codes
used in practice, wholesale algorithmic changes, such as changing the spatial
discretization or the governing equations themselves, are often highly invasive
and require a costly software re-write. Establishing a detailed and
predictive performance model for the various algorithmic choices is therefore
imperative when designing the next-generation of industry scale codes.

We establish a theoretical performance model for explicit wave-equation solvers
as used in full waveform inversion (FWI) and reverse time migration (RTM). We focus on a set of widely used
equations and establish lower bounds on the degree of the spatial
discretization required to achieve optimal hardware utilization on a set of
well known modern computer architectures. Our theoretical prediction of
performance limitations may then be used to inform algorithmic choice of future
implementations and provides an absolute measure of realizable performance
against which implementations may be compared to demonstrate their
computational efficiency.

For the purpose of this paper we will only consider explicit time stepping
algorithms based on a second order time discretization. Extension to higher
order time stepping scheme will be briefly discussed at the end. The reason we
only consider explicit time stepping is that it does not involve any matrix
inversion, but only scalar product and additions making the theoretical
computation of the performance bounds possible. The performance of other
classical algorithm such as matrix vector products or FFT as described by
\citep{Patterson} has been included for illustrative purposes.

\section{Introduction to stencil computation}

\begin{figure}
\centering
    \begin{minipage}[b]{0.4\textwidth}
	\centering
        \includegraphics[width=\textwidth]{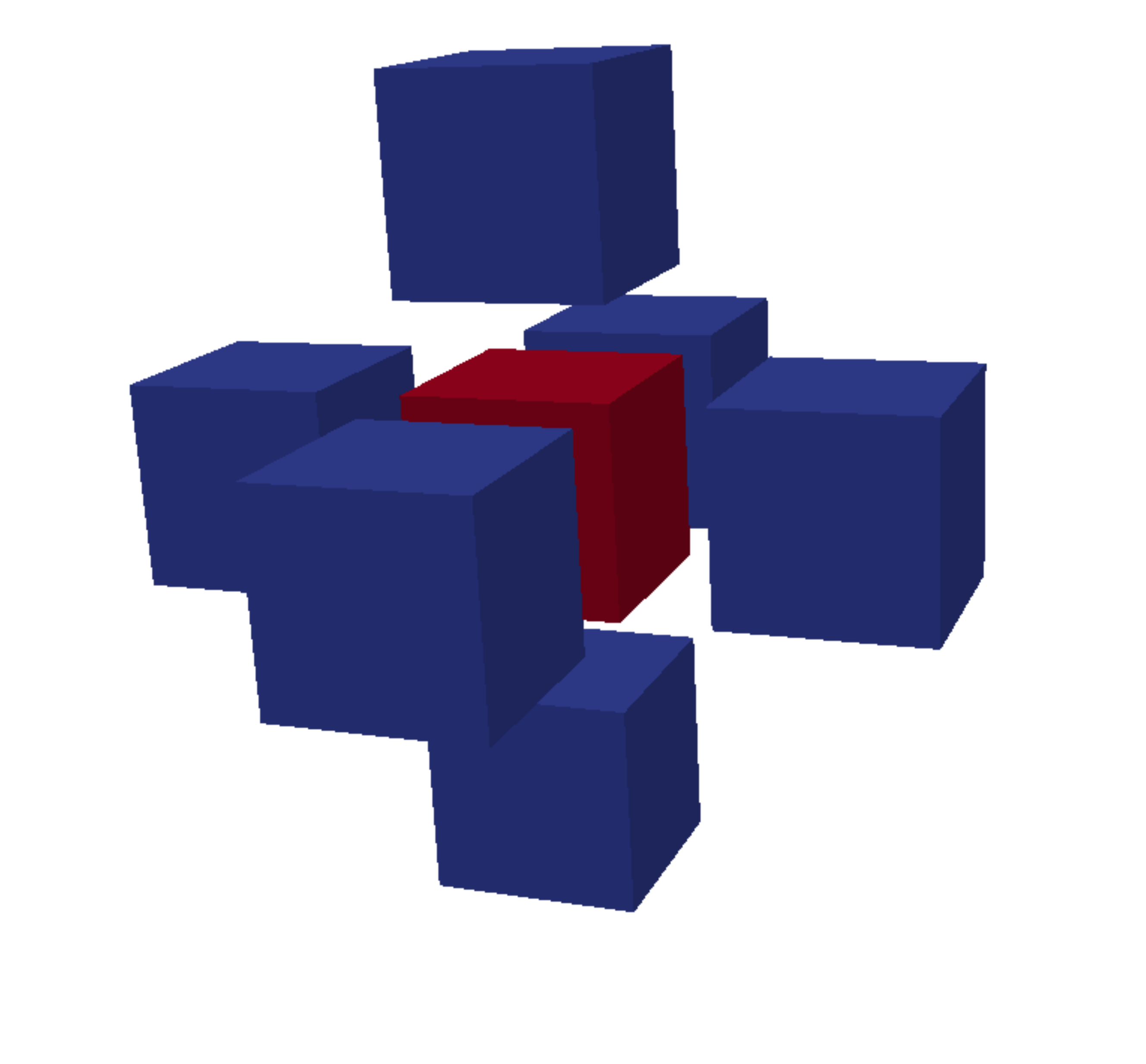}
		a)
        \label{fig:acou2}
    \end{minipage}
    \begin{minipage}[b]{0.4\textwidth}
	\centering
        \includegraphics[width=\textwidth]{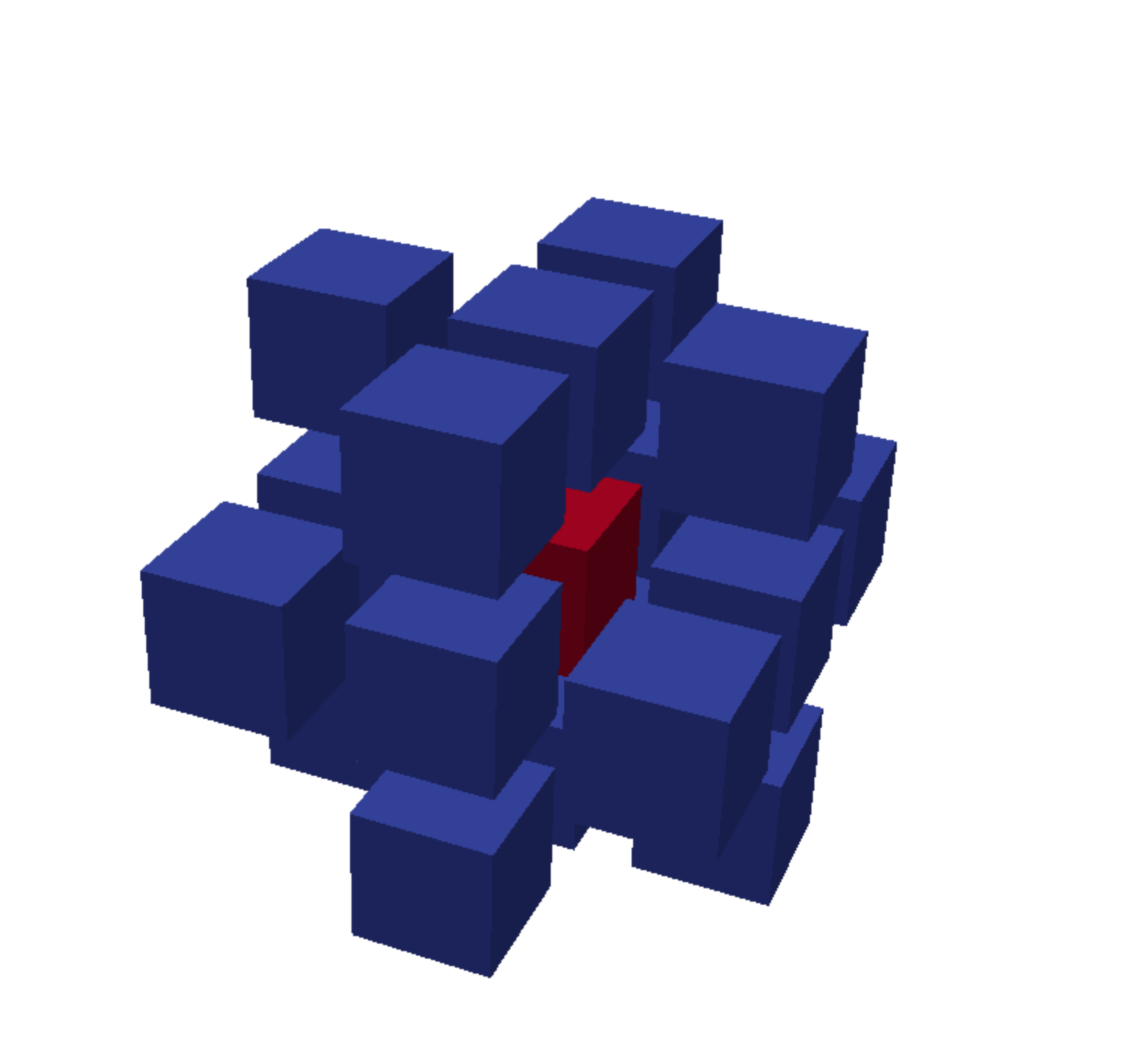}
		b)
        \label{fig:ani2}
    \end{minipage}
	\begin{minipage}[b]{0.4\textwidth}
	\centering
        \includegraphics[width=\textwidth]{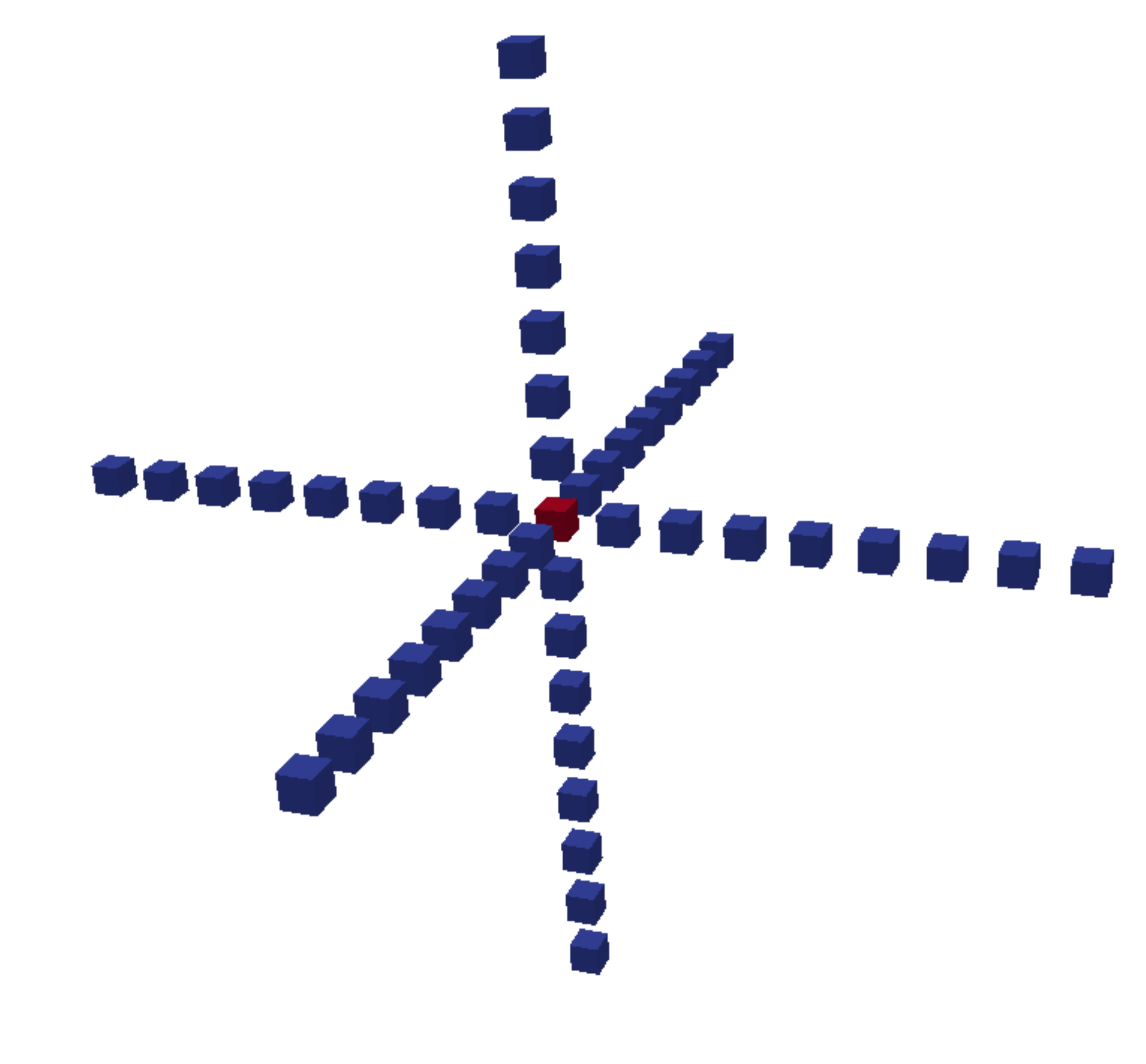}
		c)
        \label{fig:acou16}
    \end{minipage}
    \begin{minipage}[b]{0.4\textwidth}
	\centering
        \includegraphics[width=\textwidth]{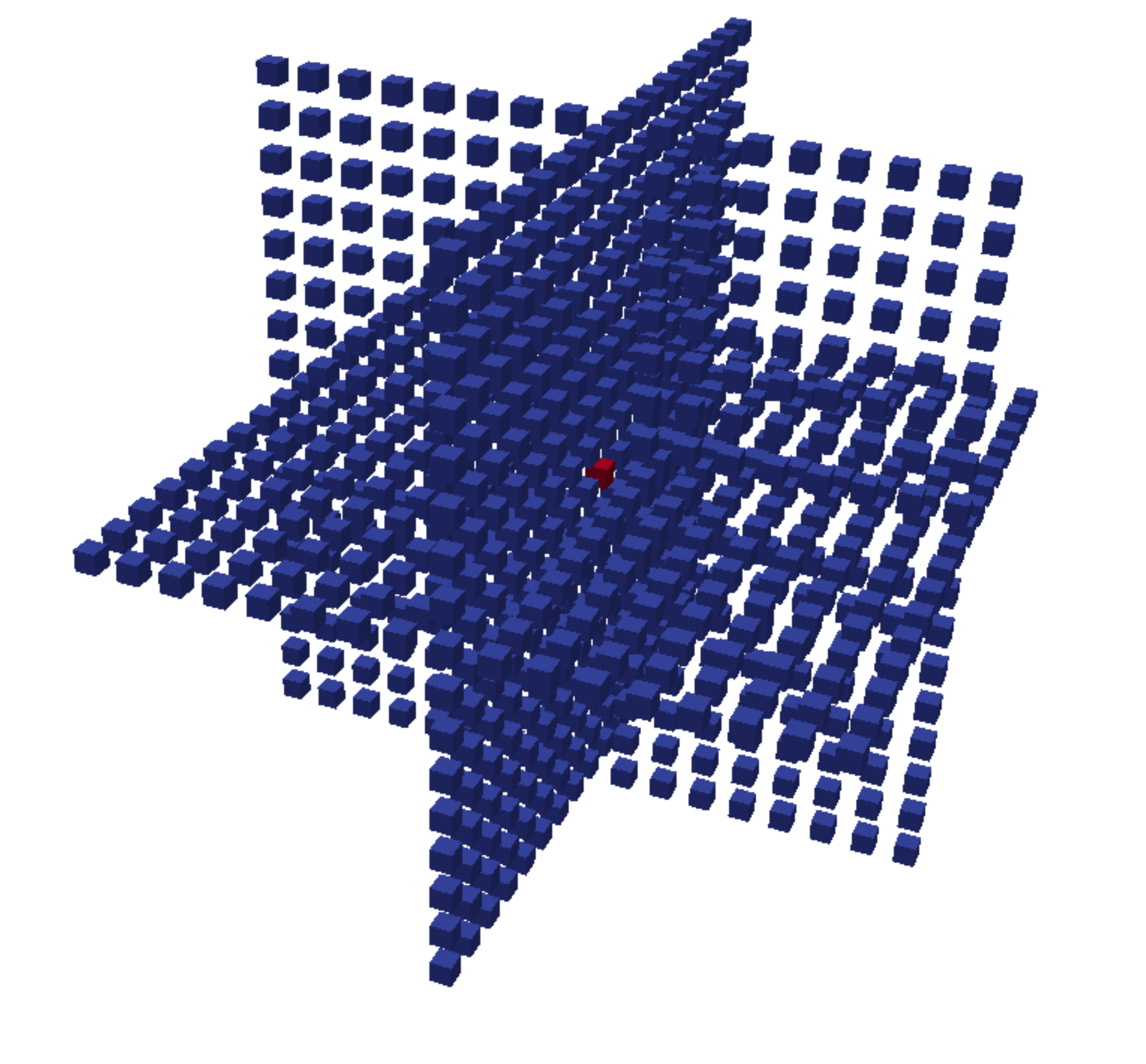}
		d)
        \label{fig:ani16}
    \end{minipage}
\caption{Stencil for the acoustic and anisotropic wave-equation for different orders of discretization. A new value for the centre point (red) is obtained by weighted sum of the values in all the neighbour points (blue). a) 2nd order laplacian, b) second order rotated Laplacian, c) 16th order Laplacian, d) 16th order rotated Laplacian}
\label{fig:stencil}
\end{figure}

\begin{figure}
\centering
	\begin{minipage}[b]{0.4\textwidth}
	\centering
        \includegraphics[width=\textwidth]{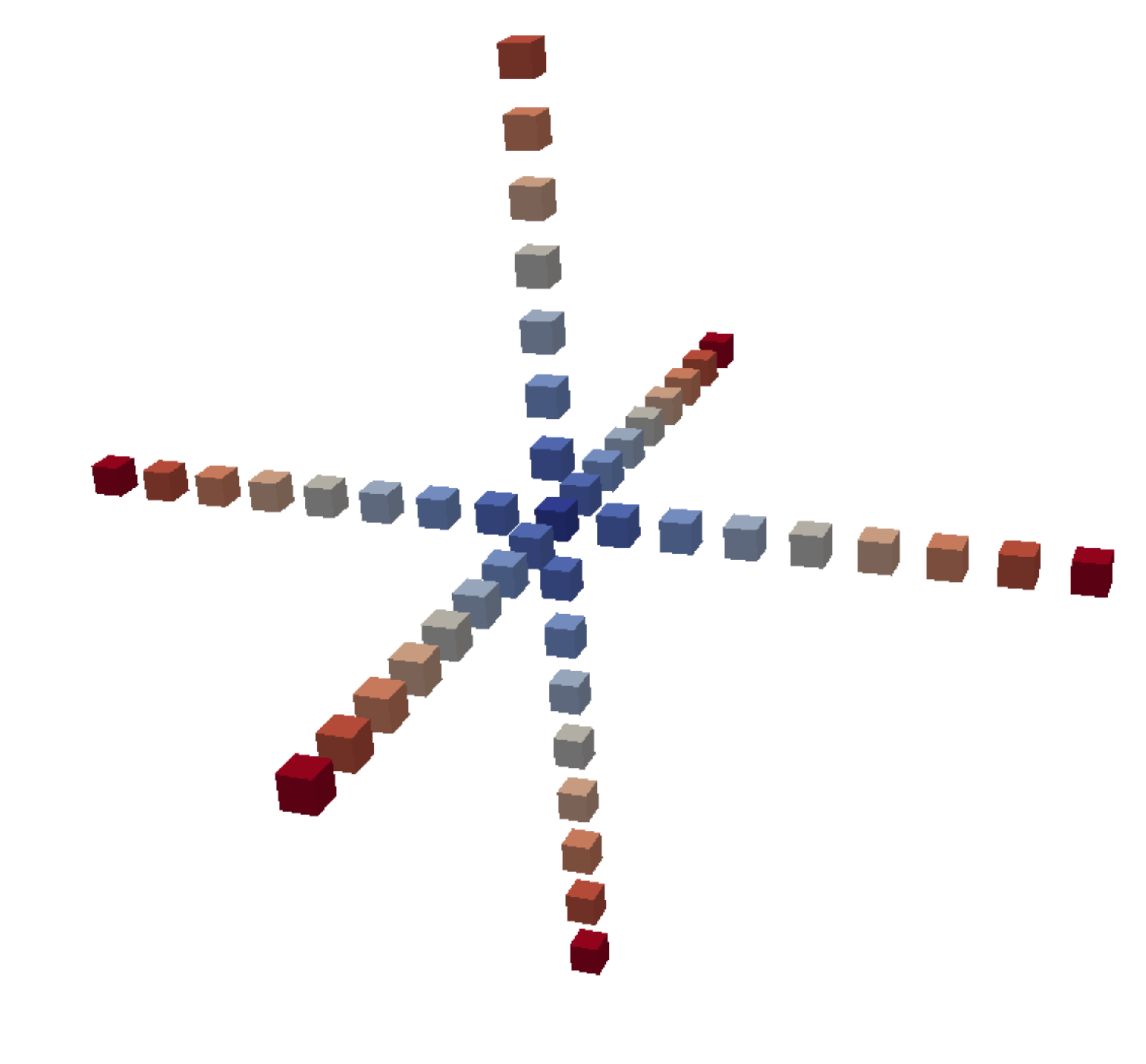}
		a)
        \label{fig:acour}
    \end{minipage}
    \begin{minipage}[b]{0.4\textwidth}
	\centering
        \includegraphics[width=\textwidth]{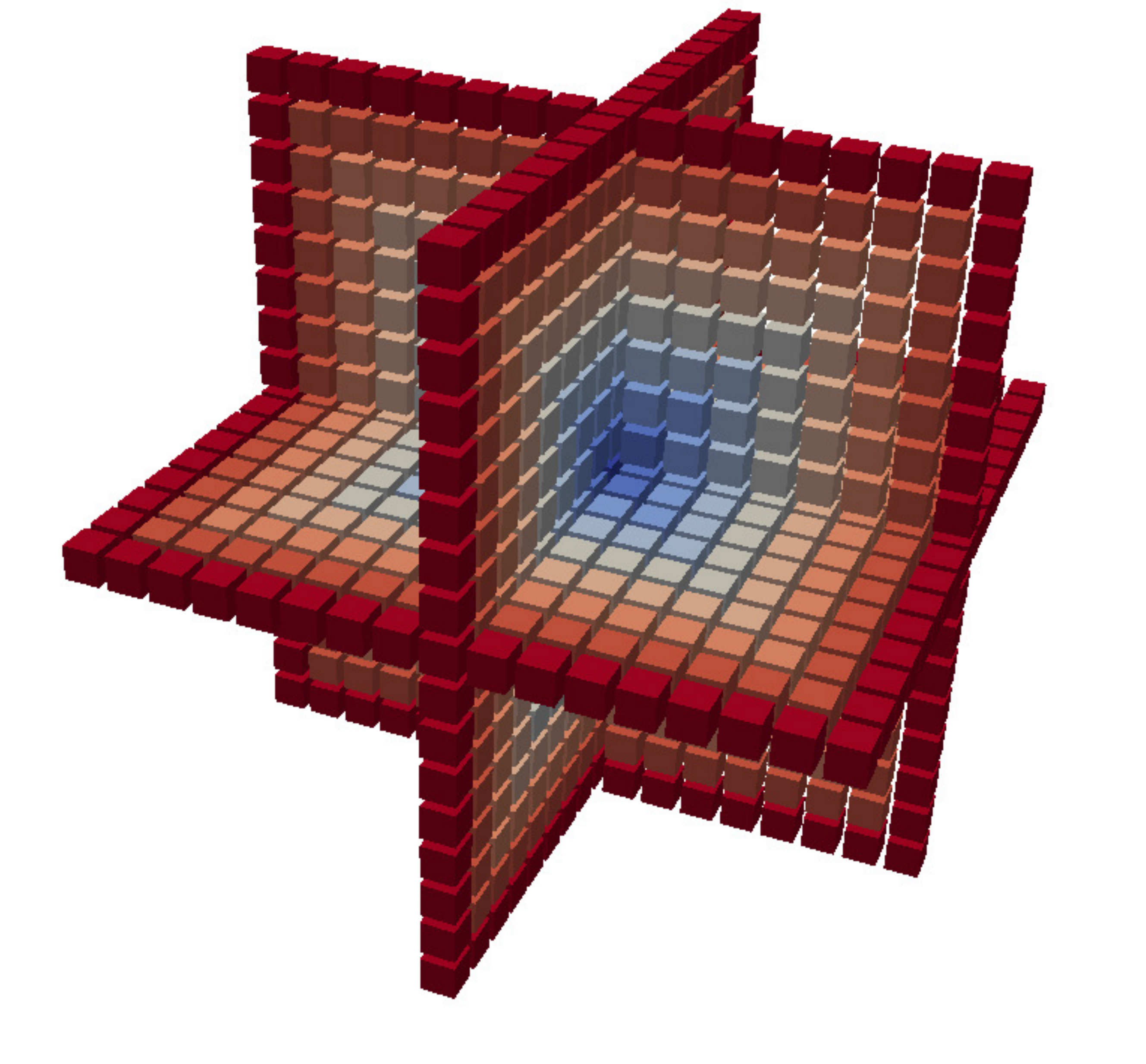}
		b)
        \label{fig:anir}
    \end{minipage}
\caption{Stencil for the 16th order acoustic and anisotropic wave-equation with distance to centre highlighting  a) Laplacian, b) rotated Laplacian}
\label{fig:stencilr}
\end{figure}

A stencil algorithm is designed to update or compute the value of a field in one spatial location according to the neighbouring ones. In the context of wave-equation solver, the stencil is defined by the support (grid locations) and the coefficients of the finite-difference scheme. We illustrate the stencil for the Laplacian, defining the stencil of the acoustic wave-equation (Eq.~\ref{eqn:Acou}), and for the rotated Laplacian used in the anisotropic wave-equation (Eq.~\ref{eqn:TTI}, ~\ref{eqn:DiffOp}) on Fig.~\ref{fig:stencil} -~\ref{fig:stencilr}. The points coloured in blue are the value loaded while the point coloured in red correspond to a written value. 


The implementation of a time stepping algorithm for a wavefield $u$, solution of the acoustic wave-equation (Eq.~\ref{eqn:Acou}) is straightforward from the representation of the stencil. We do not include the absorbing boundary conditions (ABC) as depending on the choice of implementation it will either be part of the stencil or be decoupled and treated separately.

\begin{algorithm}
\caption{Time-stepping}
\begin{algorithmic}
\FOR{$t = 0$ \TO $t=n_t$}
\FOR{$(x,y,z) \in (X,Y,Z)$}
    \STATE{$$u(t,x,y,z) = 2 u(t-1,x,y,z) - u(t-2,x,y,z) + \sum_{i \in stencil}{a_i u(t-1, x_i,y_i,z_i)}$$}
\ENDFOR
\STATE{Add Source : $ u(t,.,.,.) = u(t,.,.,.) + q$}
\ENDFOR
\end{algorithmic}
\label{alg:ts}
\end{algorithm}

In Algorithm ~\ref{alg:ts}, $(X,Y,Z)$ is the set of all grid positions in the computational domain, $(x,y,z)$ are the local indices ,$(x_i, y_i, z_i)$ are the indices of the stencil positions for the centre position $(x,y,z)$ and $n_t$ is the number of time steps and $q$ is the source term decoupled from the stencil. In the following we will concentrate on the stencil itself, as the loops in space and time do not impact the theoretical performance model we introduce. The roofline model is solely based on the amount of input/output (blue/red in the stencils) and arithmetic operations (number of sums and multiplication) required to update one grid point, and we will prove that the optimal reference performance is independent of the size of the domain (number of grid points) and of the number of time steps.

\newpage
\textbf{Notes on parallelization:}

Using a parallel framework to improve an existing code is one of the most used tool in the current stencil computation community. It is however crucial to understand that this is not an algorithmic improvement from the operational intensity. We will prove that the algorithmic efficiency of a stencil code is independent of the size of the model, and will therefore not be impacted by a domain-decomposition like parallelization via OpenMP or MPI. The results shown in the following are purely dedicated to help the design of a code from an algorithmic point of view, while parallelization will only impact the performance of the implemented code by improving the hardware usage.

\section{Roofline Performance Analysis}\label{performance-analysis}
The roofline model is a performance analysis framework designed to evaluate the
floating point performance of an algorithm by relating it to memory bandwidth
usage~\citep{williams2009roofline}. It has proved to be very popular because it
provides a readily comprehensible performance metric to interpret runtime
performance of a particular implementation according to the achievable optimal
hardware utilization for a given architecture~\citep{Williams2008}.
{\color{black}
This model has been applied to real-life codes in the past to analyze and 
report performance including oceanic climate models \cite{epicoco2014roofline}, 
combustion modeling \cite{chan2013software} and even seismic imaging 
\cite{andreolli2014genetic}. It has also been used to evaluate the effectiveness of 
implementation-time optimizations like autotuning \cite{datta2009auto}, or 
cache-blocking on specific hardware platforms like vector processors 
\cite{sato2009performance} and GPUs \cite{kim2011performance}. 
Tools are available to plot the machine-specific parameters
of the roofline model automatically \cite{RooflineModelToolkit}.
When more information about the target hardware is available, it is possible to refine 
the roofline model into the cache-aware roofline model which gives more accurate 
predictions of performance \cite{CacheAwareRoofline}. The analysis presented here 
can be extended to the cache-aware roofline model but in order to keep it general, we 
restrict it to the general roofline model. 

The roofline model has also been used to compare different types of basic numerical operations 
to predict their performance and feasibility on future systems \cite{barba2013will}, 
quite similar to this paper. However, in this paper, instead of comparing stencil 
computation to other numerical methods, we carry out a similar comparison between 
numerical implementations using different stencil sizes. This provides an upper-bound 
of performance on any hardware platform at a purely conceptual stage, long before the 
implementation of the algorithm.

Other theoretical models to predict upper-bound performance of generic code on 
hypothetical hardware have been built \cite{lai2013performance, wahib2014scalable, ECMModel, ActiveLearning} 
but being too broad in scope, can not be used to drive algorithmic choice 
like choosing the right discretization order. Some of these models have also
been applied to stencil codes \cite{ECMModelStencil, datta2009optimization}, however the analysis 
was of a specific implementation and could not be applied in general. There are many tools to perform
performance prediction at the code-level \cite{KernCraft, PerfBoundsCodeNarayanan, exasat, rahman2011understanding}.
However, any tool that predicts performance based on a code is analyzing the implementation
and not the algorithm in general. Although performance modeling is a deep and mature field, most work 
is restricted to modeling the performance of specific implementations in code. \citeauthor{KahanScalar} makes a 
comparison quite similar to the one we do here where two algorithmic choices for the same problem are 
being compared with a performance model.
}

In this section we demonstrate how one creates a roofline model for a given
computer architecture, and derives the operational intensity for a given
numerical algorithm. This establishes the theoretical upper-bound for the
performance of a specific algorithm on that architecture. A general roofline
performance analysis consists of three steps:
\begin{itemize}
\item The memory bandwidth, bytes per second, and the peak number of floating point
      operations per second (FLOPS) of the computer architecture are established
      either from the manufacturers specification or through measurement using
      standard benchmarks.
\item The operational intensity (OI) of the algorithm is established
      by calculating the ratio of the number of floating point operations
      performed to memory traffic, FLOPs per byte. This number characterizes the
      algorithmic choices that affect performance on a computer system. In
      combination with the measured memory bandwidth and peak performance
      of a computer architecture, this provides a reliable estimate of the
      maximum achievable performance.
\item The solver is benchmarked in order to establish the achieved
      performance. A roofline plot can be created to illustrate how the
      achieved performance compares to the maximum performance predicted by
      the roofline for the algorithms OI. This establishes a measure of optimality of the
	  implementation, or alternatively the maximum
      possible gain from further optimization of the software.
\end{itemize}

\subsection{Establishing the Roofline}\label{roofline-model}

The roofline model characterises a computer architecture using two parameters:
the maximum memory bandwidth, $B_{peak}$, in units of $bytes/s$; and the peak
FLOPS achievable by the hardware, $F_{peak}$. The maximally achievable
performance $F_{ac}$ is modelled as:

\begin{equation}
F_{ac} = \min\left(\mathcal{I} B_{peak}, F_{peak}\right),
\label{perf-limits}
\end{equation}
where $\mathcal{I}$ is the OI.

\begin{figure} \centering
\includegraphics[width=1.000\hsize]{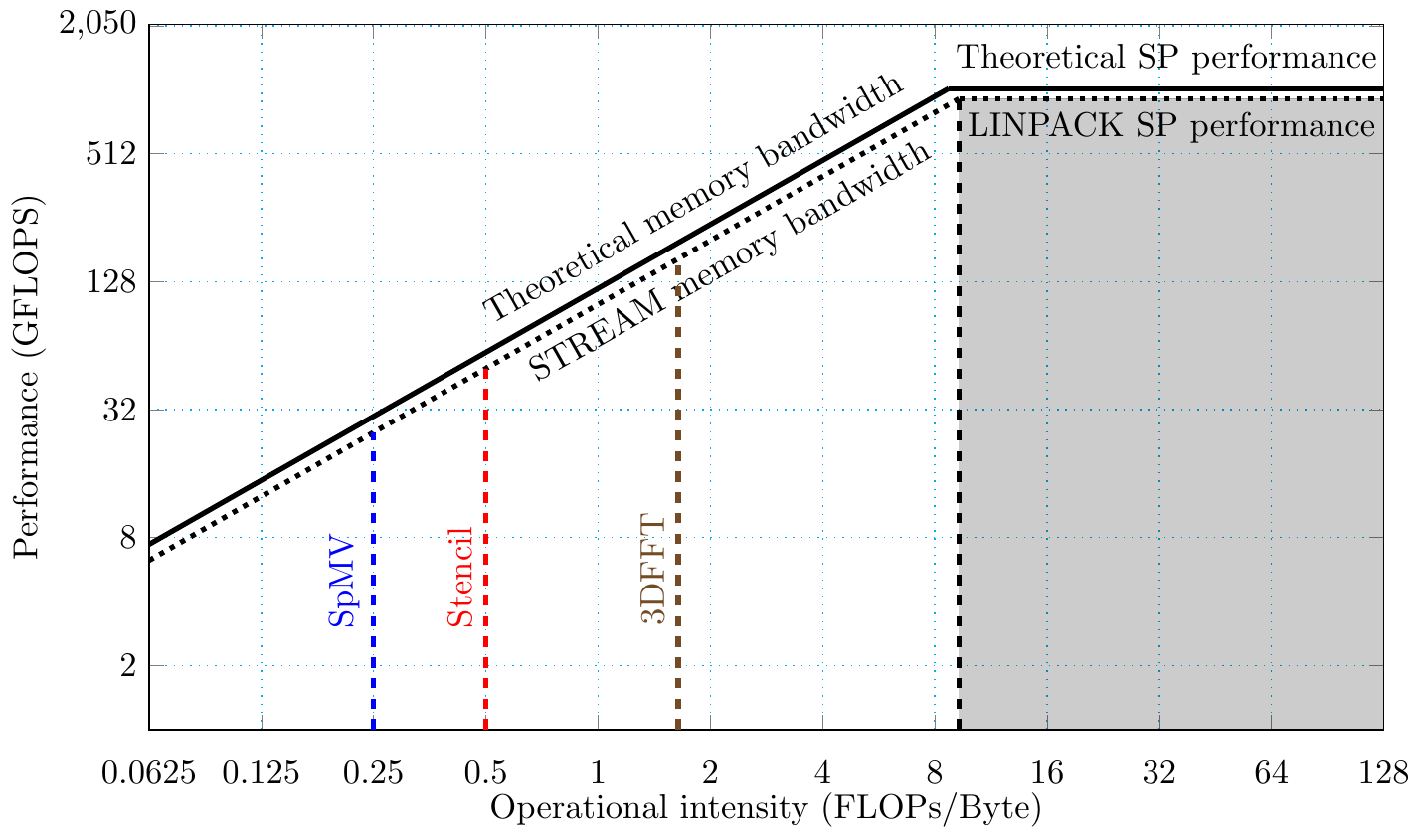}
\caption{Roofline diagram showing the operational intensity of three
well-known algorithms as reported by \citet{williams2009roofline}:
sparse matrix-vector multiplication (SpMV), stencil computation and 3D
Fast Fourier Transform (3DFFT). The hardware limits are taken
from \citet{Andreolli2015} and the compute-limited area is highlighted
through shading.}
\label{fig:ExampleRoof}
\end{figure}

As illustrated in Fig.~\ref{fig:ExampleRoof} this limitation defines two
distinct regions:
\begin{itemize}
\item
  {\bf Memory-bound}: The region left of the ridge point constitutes algorithms that
  are limited by the amount of data coming into the CPU from memory. Memory-bound
  codes typically prioritise caching optimizations, such as data reordering and cache blocking.
\item
  {\bf Compute-bound}: The region right of the ridge point
  contains algorithms that are limited by the maximum performance of
  the arithmetic units in the CPU and thus defines the maximum
  achievable performance of the given architecture. Compute-bound
  codes typically prioritise vectorization to increase throughput.
\end{itemize}

It is worth noting that changing from single to double-precision arithmetic
halves the OI because the volume of memory that must be transferred between the main memory
and the CPU is doubled. The peak performance will be impacted as well, since
the volume of data and the number of concurrently used floating point units
(FPU) changes. As commonly employed by industry, we assume single precision arithmetic for the examples presented here, but it is straightforward to {\color{black}extend} to double precision.

\citet{Andreolli2015} illustrates an example of deriving the theoretical
performance for a system that consists of two Intel Xeon E5-2697 v2 (2S-E5)
with 12 cores per CPU each running at 2.7 Ghz without turbo mode.  Since these
processors support 256-bit SIMD instructions they can process eight
single-precision operations per clock-cycle (SP FP). Further, taking into account the
use of Fused Multiply-Add (FMA) operations (two per cycle), this yields
{\color{black}
\begin{equation*}
	\begin{aligned}
F_{peak} &= 8 (SP FP) \times 2 (FMA) \times 12 (cores) \times 2 (CPUs)
\times 2.7 \text{Ghz} \\
&= 1036.8\ \text{GFLOPS.}
\end{aligned}
\end{equation*}}
Clearly, this assumes full utilization of two parallel pipelines for Add and Multiply operations. 

A similar estimate for the peak memory bandwidth $F_{peak}$ can be made from
the memory frequency ($1866 \ GHz$), the number of channels ($4$) and the number
of bytes per channel ($8$) and the number of CPUs ($2$) to give $F_{peak} =
1866 \times 4 \times 8 \times 2 = 119\ GByte / s$.

It is important to note here that there is an instruction execution overhead
that the above calculations did not take into account and therefore these
theoretical peak numbers are not achievable ($\simeq 80\%$ is achievable in practice ~\citep{Andreolli2015}). For this reason, two benchmark algorithms, STREAM TRIAD for memory bandwidth \citep{McCalpin1995,
McCalpin2007} and LINPACK for floating point performance
\citep{Dongarra:1987:LBE:647970.742568}, are often used to measure the practical
limits of a particular hardware platform. These algorithms are known to achieve
a very high percentage of the peak values and are thus indicative of practical
hardware limitations.

\subsection{Performance Model}\label{performance-model}

The key measure to using the roofline analysis as a guiding tool for
algorithmic design decisions and implementation optimization is the operational
intensity, $\mathcal{I}$, as it relates the number of FLOPs to the number of
bytes moved to and from RAM. $\mathcal{I}$ clearly does not capture many
important details about the implementation such as numerical accuracy or time
to solution. Therefore, it is imperative to look at $\mathcal{I}$ in
combination with these measures when making algorithmic choices.  

Here we analyze the algorithmic bounds of a set of finite-difference
discretizations of the wave-equation using different stencils and spatial
orders. We therefore define algorithmic operational intensity
$\mathcal{I}_{alg}$ in terms of the total number of FLOPs required to compute a
solution, and we assume that our hypothetical system has a cache with infinite
size and no latency inducing zero redundancy in memory traffic \citep{Williams2008}. This acts as a
theoretical upper bound for the performance of any conceivable implementation.  

We furthermore limit our theoretical investigation to analysing a single time
step as an indicator of overall achievable performance. This assumption allows
us to generalize the total number of bytes in terms of the number of spatially dependant variables
(e.g. wavefields, physical properties) used in the discretized equation as
$\mathcal{B}_{global} = 4 N (l + 2 s)$, where $l$ is the number of variables
whose value is being loaded, $s$ is the number of variables whose value is
being stored, $N$ is the number of grid points and $4$ is the number of bytes
per single-precision floating point value. The term $2 s$ arises from the fact
that most computer architectures will load a cache line before it gets
overwritten completely. However, some computer architectures, such as the Intel
Xeon Phi, have support for stream stores, so that values can be written
directly to memory without first loading the associated cache line, in which
case the expression for the total data movement becomes $\mathcal{B}_{global}
= 4 N (l + s)$. It is important to note here that limiting the analysis to a
single time step limits the scope of the infinite caching assumption above.

Since we have assumed a constant grid size $N$ across all spatially dependant
variables, we can now parametrize the number of FLOPs to be computed per time
step as $\mathcal{F}_{total}(k) = N \mathcal{F}_{kernel}(k)$, where
$\mathcal{F}_{kernel}(k)$ is a function that defines the number of flops
performed to update one grid point in terms of the stencil size $k$ used to
discretize spatial derivatives.  Additional terms can be added corresponding to
source terms and boundary conditions but they are a small proportion of the
time step in general and are neglected here for simplicity. This gives us the
following expression for OI as a function of $k$, $\mathcal{I}_{alg}(k)$:

\begin{equation}
\mathcal{I}_{alg}(k) = \mathcal{F}_{total}(k) / \mathcal{B}_{global} = \frac{\mathcal{F}_{kernel}(k)}{4(l+s)}.
\label{eqn_oi}
\end{equation}

\section{Operational intensity for finite-differences}
\label{the-roofline-model-for-pde-solvers}

We derive a general model for the operational intensity of
wave-equation PDEs solvers with finite-difference discretizations
using explicit time stepping and apply it to three different wave-equation
formulations commonly used in the oil and gas exploration community: an acoustic anisotropic wave-equation; vertical transverse isotropic (VTI); and tilted transversely isotropic (TTI) ~\citep{liu2009stable}. The theoretical
operational intensity for the 3D discretized equations will be calculated as a function
of the finite-difference stencil size $k$, which allows us to make predictions about the minimum
discretization order required for each algorithm to reach the compute-bound regime for
a target computer architecture. For completeness we describe the equations in
Appendix ~\ref{PDE}.

\subsection{Stencil operators}\label{stencil-operators}

As a baseline for the finite-difference discretization, we consider the use of a 1D symmetric stencil of size $k$,
which uses $k$ values of the discretized variable to compute any spatial derivatives enforcing a fixed support for all derivatives. Other choices of discretization are possible, such as choosing the stencil for the first derivative and applying it iteratively to obtain high order derivatives. Our analysis will still be valid but require a rewrite of the following atomic operation count. The number of FLOPs used for the three types of derivatives involved in our equation are calculated as:

\begin{itemize}
\item first order derivative with respect to $x_i$ ($\frac{du}{dx_i}$):
      $(k + 1)\textrm{ mult } + (k - 1)\textrm{ add }= 2k\ FLOPs$
\item second order derivative with respect to $x_i$ ($\frac{d^2u}{dx^2_i}$):
      $(k + 1)\textrm{ mult } + (k - 1)\textrm{ add } = 2k\ FLOPs$
\item second order cross derivative with respect to $x_i, x_j$ ($\frac{d^2u}{dx_i dx_j}$):
      $(k^2 - 2k)\textrm{ mult } + (k^2 - 2k - 1)\textrm{ add } = 2k^2 -4k -1\ FLOPs$
\end{itemize}

where in 3D, $x_i \text{ for } i = 1,2,3$ correspond to the three dimensions $x,y,z$ and $u$ is the discretized field.

\begin{table}[ht] \centering
\begin{tabular}{ccccccc}
Equation & $\frac{du}{dx_i}$ & $\frac{d^2u}{dx^2_i}$ &
           $\frac{d^2u}{dx_i dx_j}$ & mult & add & duplicates \\
\hline
Acoustic:       &  0  &  $3\times2k$ &  0  &  3  &  5  & $- 4$  \\
VTI: $2\times($ &  0  &  $3\times2k$ &  0  &  5  &  5  & $- 2)$ \\
TTI: $2\times($ &  0  &  $3\times2k$ &  $3\times( 2k^2 - 4k -1)$  &  44 &  17 & $- 8)$ \\
\end{tabular}
\label{tab:flops-3d}
\caption{Derivation of $FLOPs$ per stencil invocation for each equation.}
\end{table}

\begin{figure}
\centering
\includegraphics[width=0.9500\hsize]{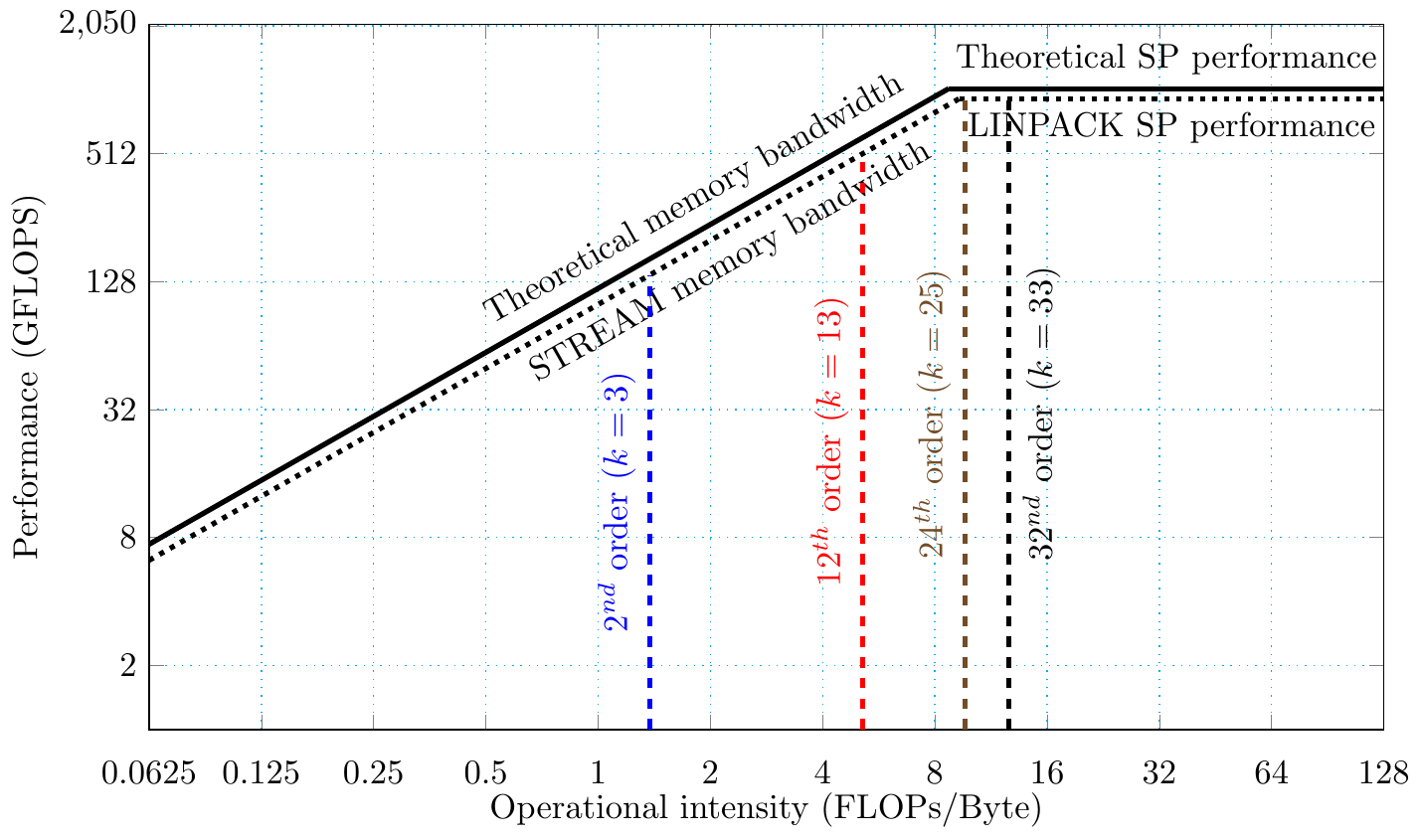}
\caption{Increase in algorithmic OI with increasing stencil sizes on a dual-socket
Intel Xeon E5-2697 v2~\citep{Andreolli2015} for a 3D acoustic kernel. The $24^{th}$ order stencil is coincident with the ridge point --- the transition point from memory-bound to compute-bound computation.}
\label{fig:RoofEqns_xeonA}
\end{figure}

\begin{figure}
\centering
\includegraphics[width=0.9500\hsize]{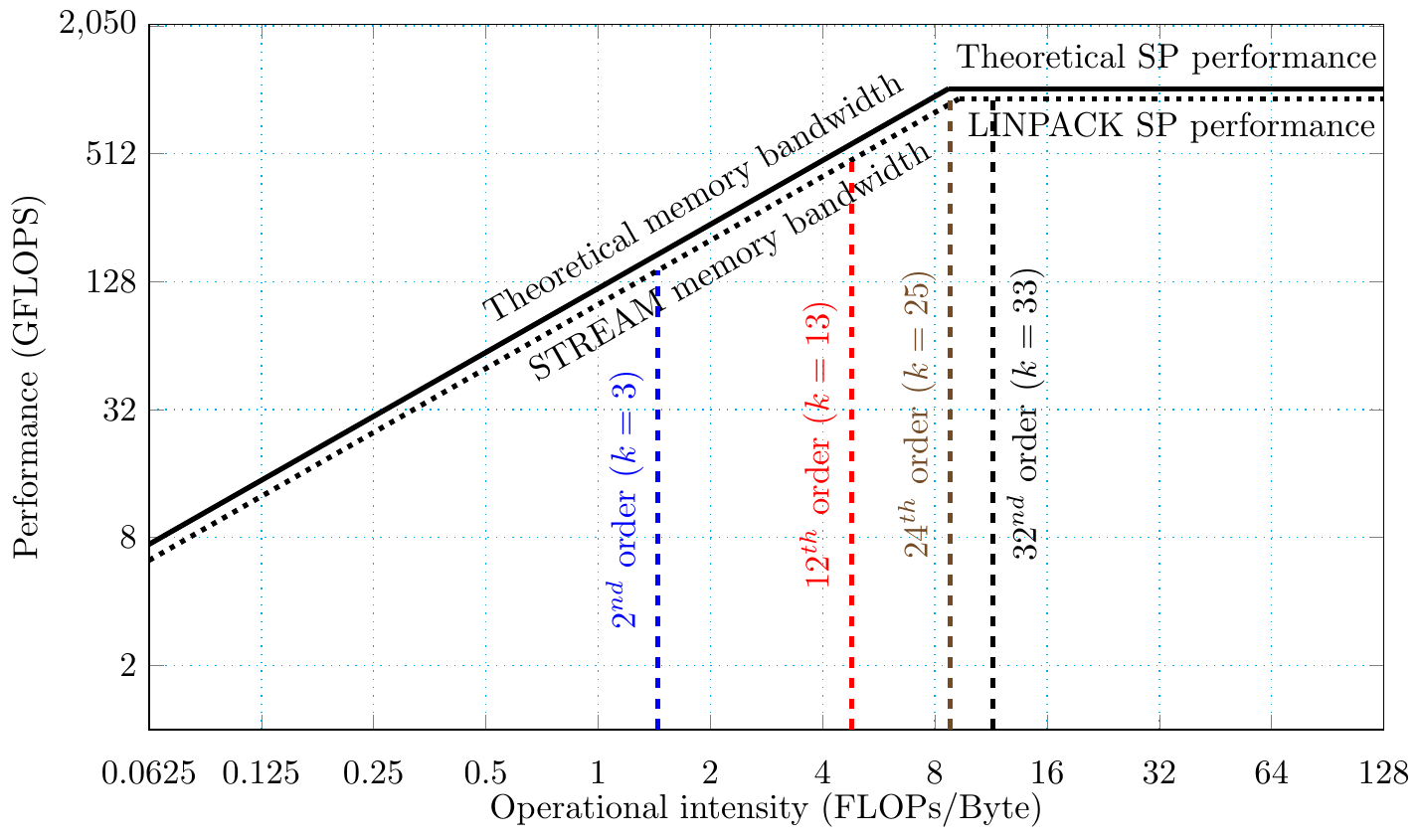}
\caption{Increase in algorithmic OI with increasing stencil sizes on a dual-socket
Intel Xeon E5-2697 v2~\citep{Andreolli2015} for a 3D VTI kernel. Similarly to the acoustic model, the $24^{th}$ order stencil is coincident with the ridge point.}
\label{fig:RoofEqns_xeonV}
\end{figure}

\begin{figure}
\centering
\includegraphics[width=0.9500\hsize]{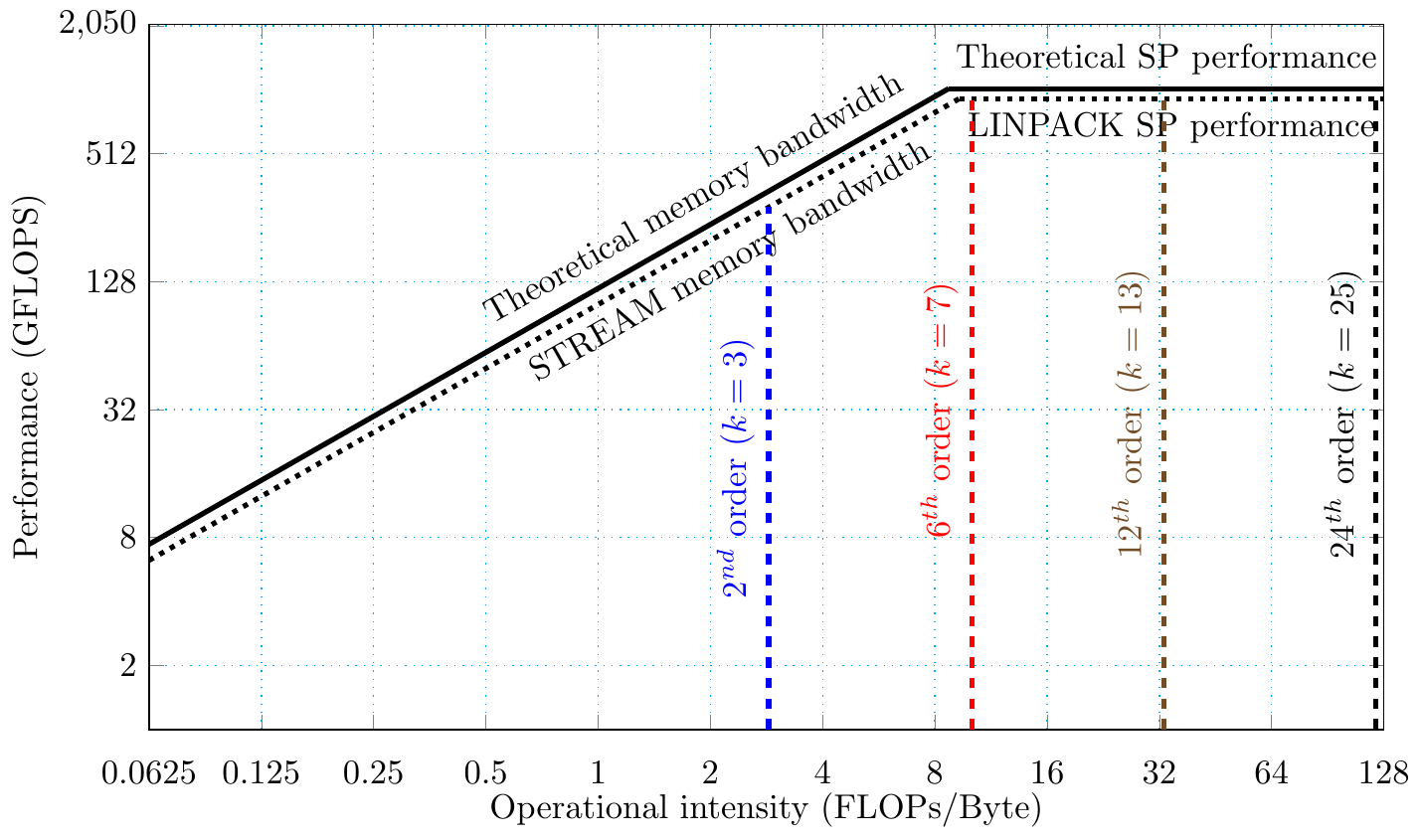}
\caption{Increase in algorithmic OI with increasing stencil sizes on a dual-socket
Intel Xeon E5-2697 v2~\citep{Andreolli2015} for a 3D TTI kernel. The $6^{th}$ order stencil is already compute-bound.}
\label{fig:RoofEqns_xeonT}
\end{figure}

\begin{figure}
\centering
\includegraphics[width=0.9500\hsize]{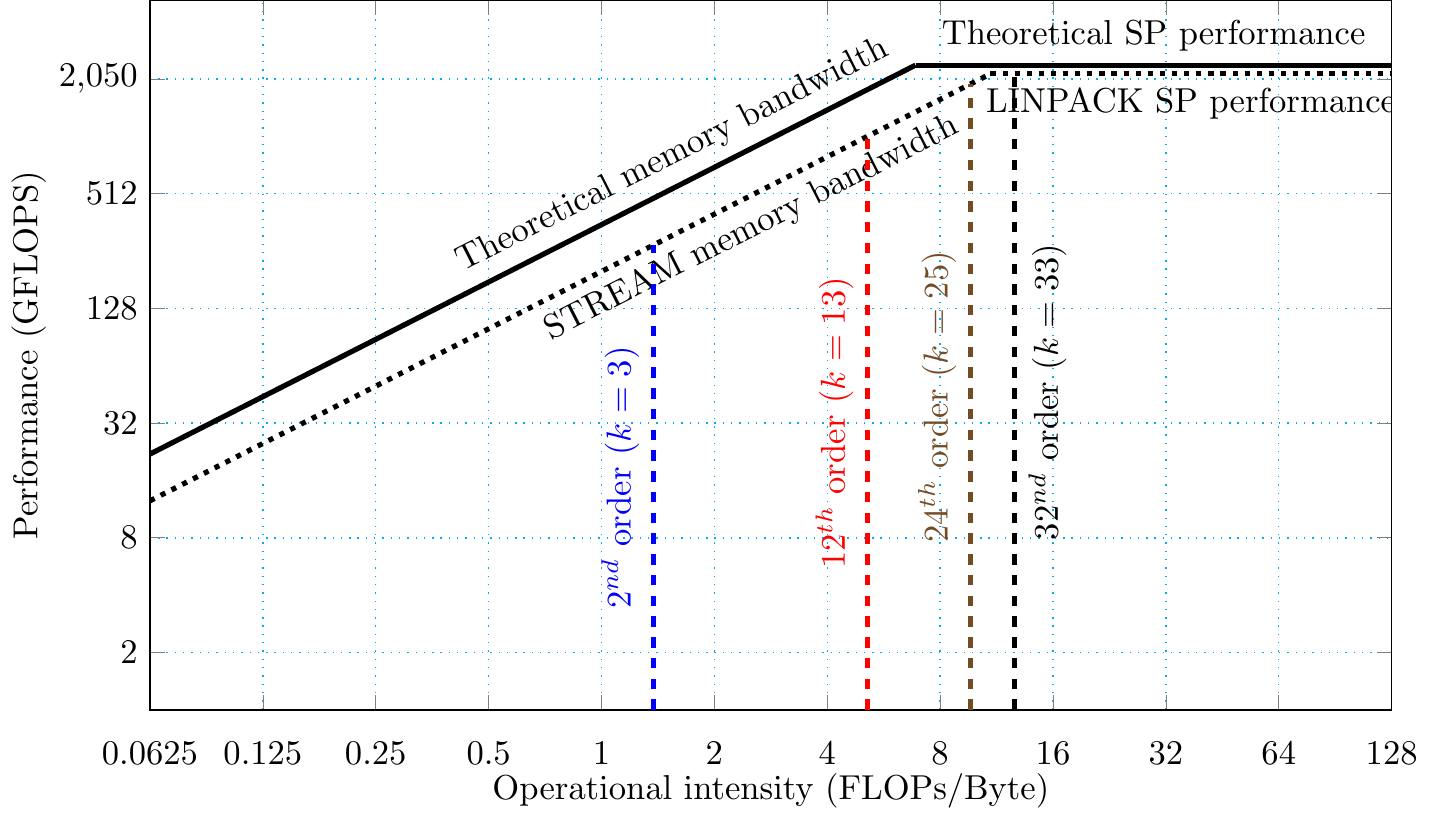}
\caption{Increase in algorithmic OI with increasing stencil sizes on a 
Intel Xeon Phi 7120A co-processor~\citep{Andreolli2015} for a 3D acoustic kernel. Unlike the Xeon E5-2697, the $30^{th}$ order stencil is the smallest one to be compute-bound (vs $24^{th}$ order).}
\label{fig:RoofEqns_phiA}
\end{figure}

\begin{figure}
\centering
\includegraphics[width=0.9500\hsize]{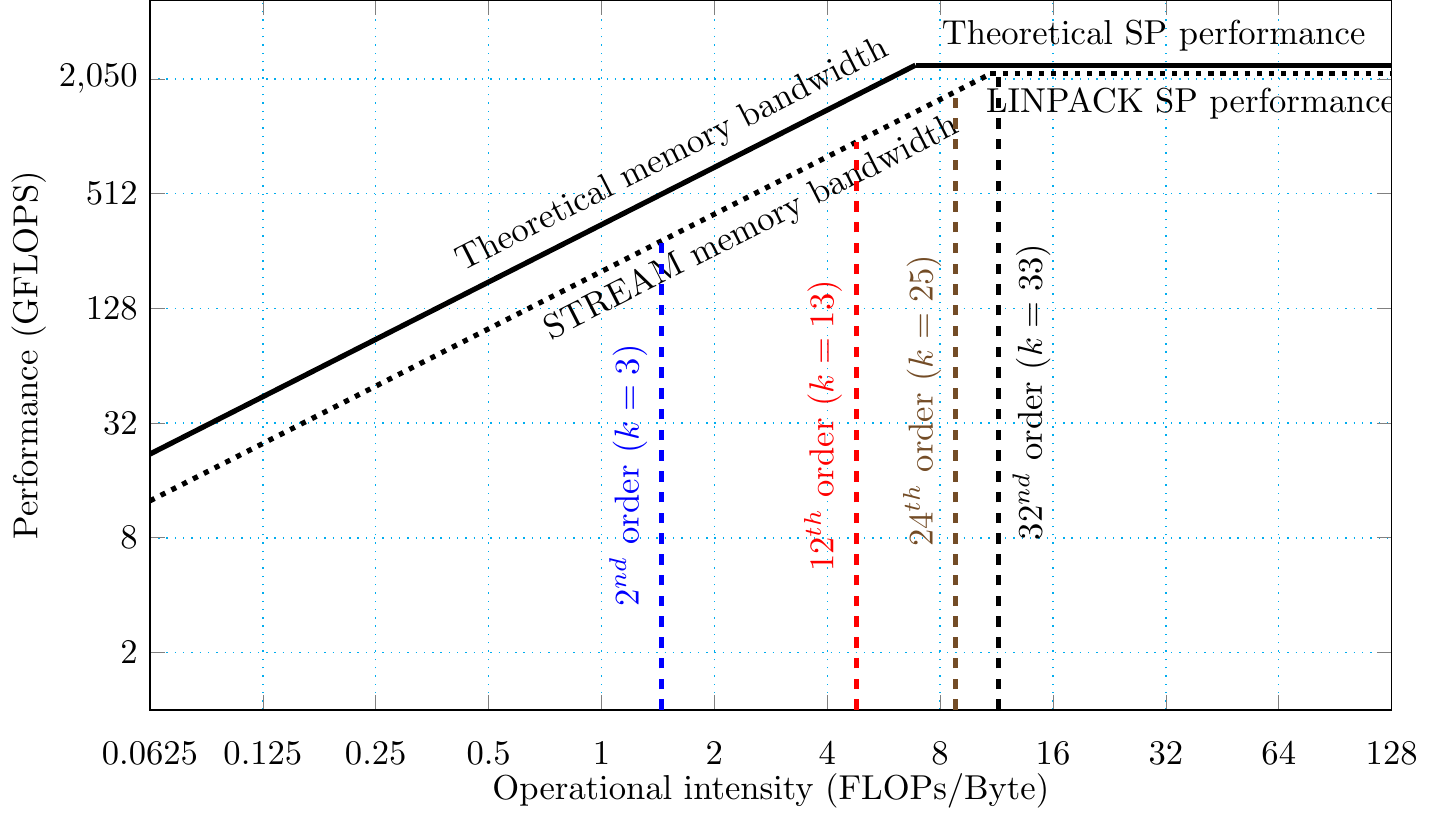}
\caption{Increase in algorithmic OI with increasing stencil sizes on a 
Intel Xeon Phi 7120A co-processor~\citep{Andreolli2015} for a 3D VTI kernel. $32^{nd}$ is the minimum compute-bound stencil. It is not equivalent to the acoustic on this architecture.}
\label{fig:RoofEqns_phiV}
\end{figure}

\begin{figure}
\centering
\includegraphics[width=0.9500\hsize]{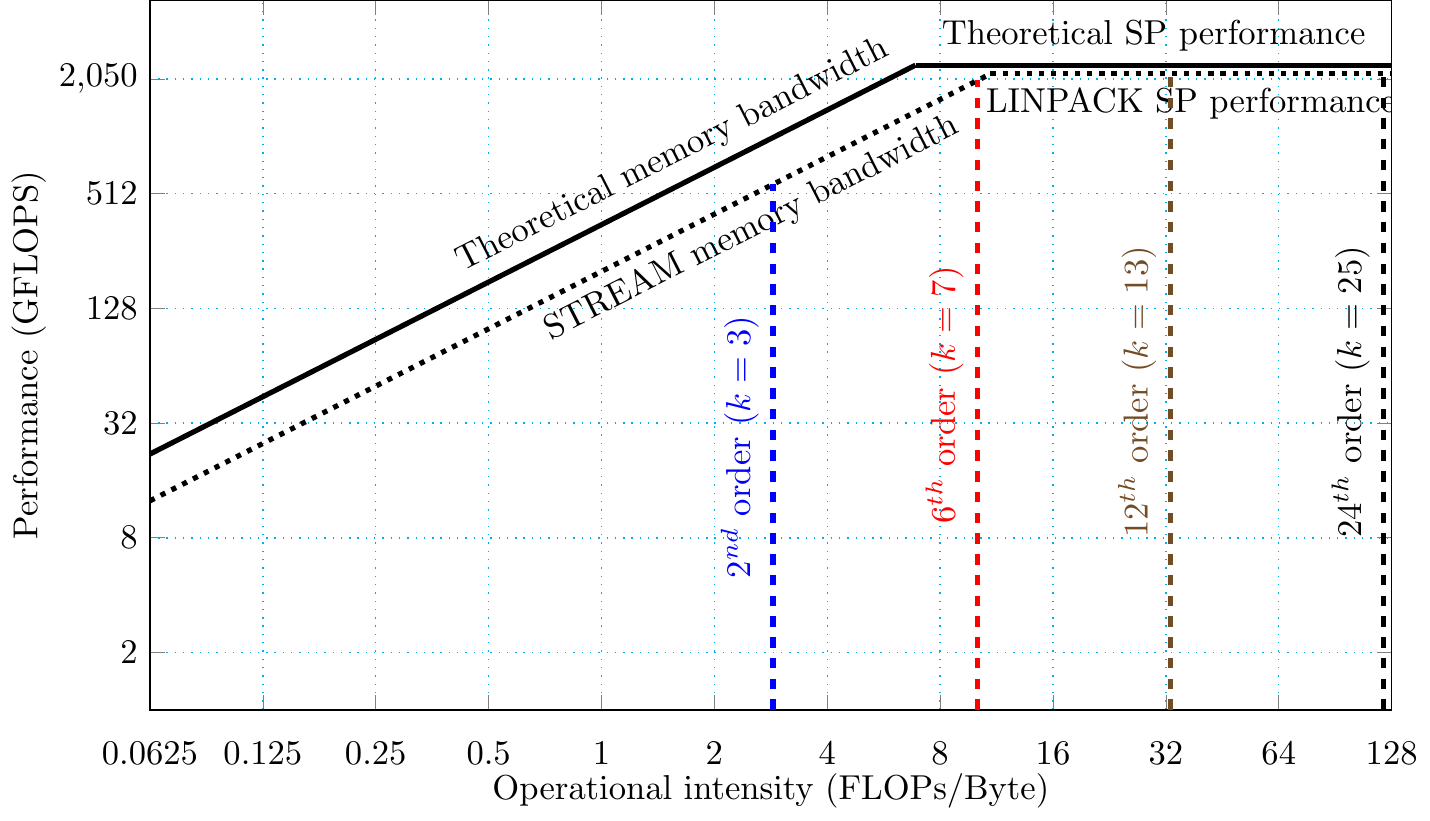}
\caption{Increase in algorithmic OI with increasing stencil sizes on a 
Intel Xeon Phi 7120A co-processor~\citep{Andreolli2015} for a 3D TTI kernel. The $6^{th}$ order stencil is already compute-bound similarly to the Xeon E5-2697.}
\label{fig:RoofEqns_phiT}
\end{figure}

\begin{figure} \centering
\includegraphics[width=0.7500\hsize]{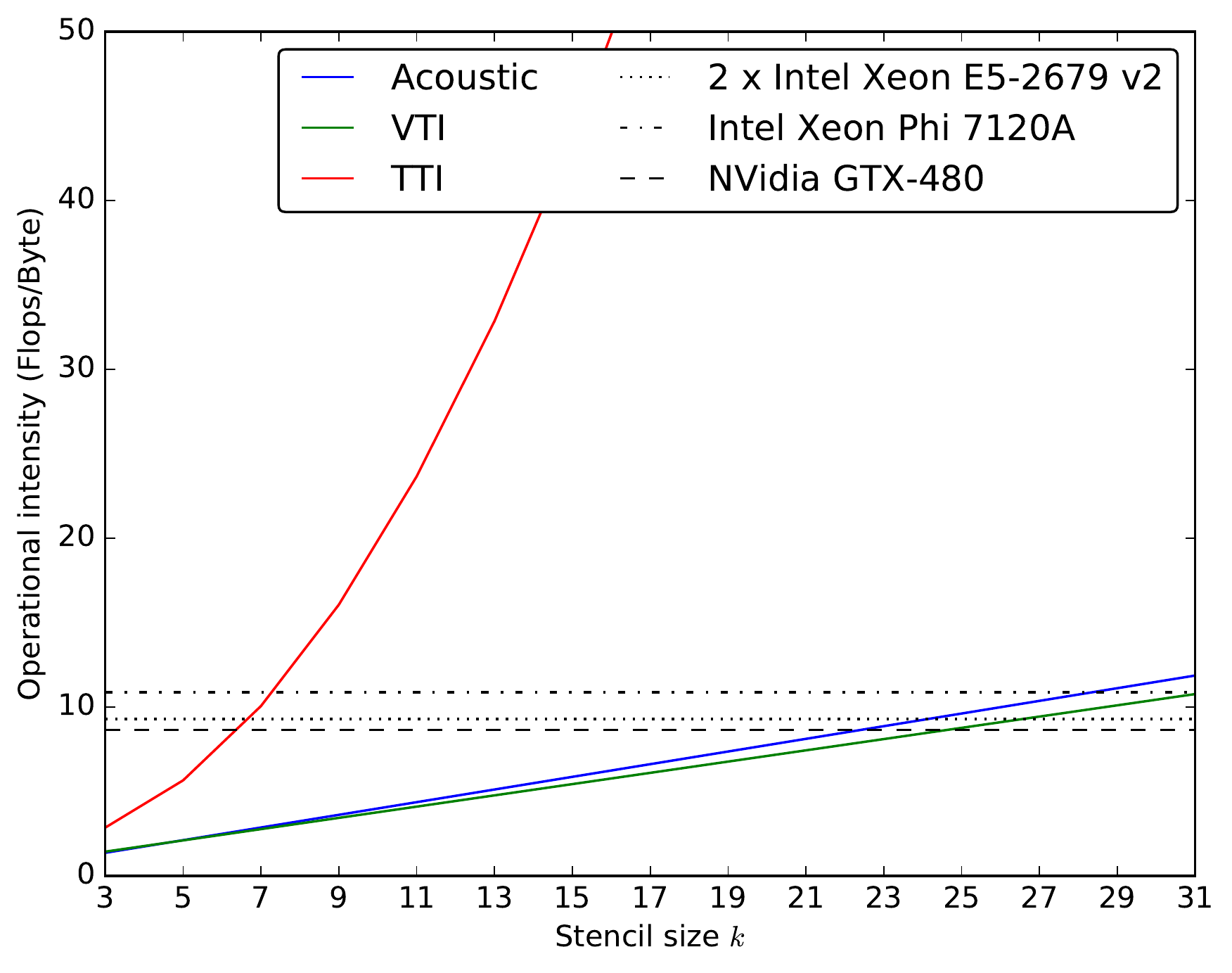}
\caption{Increase in OI with stencil size $k$ and machine-specific minimum
OI values for all three hardware architectures considered in this paper.}
\label{fig:OI_stencil_size}
\end{figure}

Computing the total wavefield memory volume $B_{global}$ for each equation we
have $4\times4N\ bytes$ for Acoustic (load velocity, two previous time steps and write the new time step), $9\times4N\ bytes$ for VTI (load velocity, two anisotropy parameters, two previous time steps for two wavefields and write the new time step for the two wavefields) and $15\times4N\ bytes$ for TTI (VTI plus 6 precomputed cos/sin of the tilt and dip angles). Eq.~\ref{eqn_oi} allows us to predict the increase of the operational intensity in terms of $k$ by replacing $B_{global}$ by its value. The OI $\mathcal{I}_{alg}(k)$ for the three wave-equations is given by:

\begin{itemize} 
\item Acoustic anisotropic: $\mathcal{I}_{alg}(k) = \frac{3k}{8} + \frac{1}{4}$,
\item VTI: $\mathcal{I}_{alg}(k) = \frac{k}{3} + \frac{4}{9}$,
\item TTI: $\mathcal{I}_{alg}(k) = \frac{k^2}{5} - \frac{k}{5} + \frac{5}{3}$,
\end{itemize}

and plotted as a function of $k$ on Fig.~\ref{fig:OI_stencil_size}. 
Using the derived formula for the algorithmic operational intensity in terms of
stencil size, we can now analyze the optimal performance for each equation
with respect to a specific computer architecture. We are using the theoretical
and measured hardware limitations reported by \citet{Andreolli2015} to
demonstrate how the main algorithmic limitation shifts from being
bandwidth-bound at low $k$ to compute-bound at high $k$
on a dual-socket Intel Xeon in Fig.~\ref{fig:RoofEqns_xeonA} - ~\ref{fig:RoofEqns_xeonT} and an Intel Xeon
Phi in Fig.~\ref{fig:RoofEqns_phiA} - ~\ref{fig:RoofEqns_phiT}.

It is of particular interest to note from Fig.~\ref{fig:RoofEqns_xeonA} that a
$24^{th}$ order stencil with $k=25$ provides just enough arithmetic load for
the 3D acoustic equation solver to become compute-bound, while $k=25$ falls
just short of the compute-bound region for the VTI algorithm. On the other hand
a $6^{th}$ order stencil with $k=7$ is enough for the TTI algorithm to become
compute-bound due to having a quadratic slope with respect to $k$ (Fig.~\ref{fig:OI_stencil_size}) instead of a linear slope.

At this point, we can define $\mathcal{I}_{min}$, which is the minimum OI
required for an algorithm to become compute-bound on a particular architecture, as the x-axis coordinate of the ridge point in Fig.~\ref{fig:RoofEqns_xeonA} -~\ref{fig:RoofEqns_xeonT} and
~\ref{fig:RoofEqns_phiA} -~\ref{fig:RoofEqns_phiT}. Note that the ridge point x-axis position changes between the two different architectures. This difference in compute-bound limit shows that a different spatial order discretization should be used on the two architecture to optimize hardware usage. As reported by \citet{Andreolli2015} the $\mathcal{I}_{min}$ as
derived from achievable peak rates is $9.3\ FLOPs/byte$ for the Intel Xeon and
$10.89\ FLOPs/byte$ for the Intel Xeon Phi. This entails that while the
acoustic anisotropic wave-equation and VTI are memory bound for discretizations
up to $24^{th}$ order, the TTI equation reaches the compute bound region with
even a $6^{th}$ order discretization.

From the analytical expression derived we can now generalize the derivation of
minimum OI values by plotting the simplified expressions for
$\mathcal{I}_{alg}(k)$ against known hardware OI limitations, as shown in
Fig.~\ref{fig:OI_stencil_size}. We obtain a theoretical prediction about the
minimum spatial order required for each algorithm to provide enough arithmetic
load to allow implementations to become compute-bound. Most importantly,
Fig.~\ref{fig:OI_stencil_size} shows that the TTI wave-equation has a significantly steeper
slope of $\mathcal{I}(k)$, which indicates that it will saturate a given
hardware for a much smaller spatial discretization than the acoustic wave or
the VTI algorithm.

Moreover, assuming a spatial discretization order of $k-1$, we can predict that
on the Intel Xeon CPU we require a minimum order of $24$ for the acoustic wave
solver, $26$ for VTI and $6$ for TTI. On the Nvidia GPU, with a slightly lower
hardware $\mathcal{I}$, we require a minimum order of $22$ for the acoustic wave solver,
$24$ for VTI and $6$ for TTI, while even larger stencils are required for the
Intel Xeon Phi accelerator: a minimum order of $28$ for the acoustic wave
solver, $30$ for VTI and $6$ for TTI. This derivation demonstrates that
overall very large stencils are required for the acoustic anisotropic solver
and VTI to fully utilize modern HPC hardware, and that even TTI requires at
least order $6$ to be able to computationally saturate HPC architectures with
a very high arithmetic throughput, like the Intel Xeon Phi.

\section{Example: MADAGASCAR modelling kernel}\label{rsf-modelling-kernel}

We demonstrate our proposed performance model and its flexibility by applying it on a broadly used and benchmarked
modelling kernel contained in Madagascar \citep{Madagascar}. We are illustrating the ease to extend our method to a different wave-equation and by extension to any PDE solver. The code implements the 3D
anisotropic elastic wave-equation and is described in \citep{Weiss2013}. We are
performing our analysis based on the space order, hardware and runtime described in
\citep{Weiss2013}. The governing equation considered is:
	
\begin{equation}
\begin{aligned}
   &\rho \frac{d^2 u_i}{dt^2} = \frac{d \sigma_{ij}}{dx_j} + F_i, \\
   &\sigma_{ij} = c_{ijkl}\epsilon_{kl}, \\
   &\epsilon_{kl} = \frac{1}{2}[\frac{u_l}{dx_k} + \frac{u_k}{dx_l}],\\
   &u_i(.,0) = 0, \\
   &\frac{d u_i(x,t)}{dt}|_{t=0} = 0.\\
\end{aligned}
\label{PDErsf}
\end{equation}

where $\rho$ is the density, $u_i$ is the $i^{th}$ component of the three dimensional wavefield displacement ($i=1,2,3$ for $x,y,z$), $F$ is the source term,$\epsilon$ is the strain tensor,
$\sigma$ is the stress tensor and $c$ is the stiffness tensor.  The
equation is discretized with an $8^{th}$ order star stencil for the first
order derivatives and a second order scheme in time and solves for all three
components of $u$. Eq.~\ref{PDErsf} uses Einstein notations
meaning repeated indices represent summation:

\begin{equation}
\begin{aligned}
    \frac{d \sigma_{ij}}{dx_j}  &= \sum_{j=1}^3 \frac{\sigma_{ij}}{dx_j},  \\
     c_{ijkl}\epsilon_{kl} &= \sum_{k=1}^3 \left( \sum_{l=1}^3 c_{ijkl}\epsilon_{kl} \right).
\end{aligned}
\label{Einstein}
\end{equation}

From this equation and knowing the finite-difference scheme used we can already
compute the minimum required bandwidth and operational intensity. We need to
solve this equation for all three components of the wave $u$ at once as we
have coupled equations in $\epsilon$ and $u$. For a global estimate of the overall memory traffic,
we need to account for loading and storing $2 \times 3N$ values of the
displacement vector and loading $N$ values of $\rho$. In case the stiffness
tensor is constant in space the contribution of $c_{ijkl}$ is $64$
independently of $N$, which yields an overall data volume of
$\mathcal{B}_{global} = 4N(6 + 1) + 64 \simeq 28N\ Bytes$.
In the realistic physical configuration of a spatially varying stiffness
tensor, we would estimate loading $64N$ values of $c_{ijkl}$, leaving us with a
data volume of $B_{global} =4N (6 + 1 + 64) = 284N\ Bytes$. Finally we
consider symmetries in the stiffness tensor are taken into account reducing the
number of stiffness values to load to $21N$ and leading to a data volume of
$B_{global} = (6 + 1 + 21)\times4N = 112N\ Bytes$.

The number of valuable FLOPs performed to update one grid point can be estimated by: 

\begin{itemize}
\itemsep1pt\parskip0pt\parsep0pt
\item
  9 first derivatives
  ($\partial_k u_l \text{, for all } k,l = 1,2,3$) : $9 \times (8$ mult {\color{black}$+\ 7$ add$) = 135\ FLOPs$}
\item
  9 sums for $\epsilon_{kl}$ ($9 \times 9$ adds) and $9 \times 8$ mult for $\sigma_{ij}$ = $153\ FLOPs$
\item
  9 first derivatives $\partial_j\sigma_{ij}$ and 9 sums =  $144\ FLOPs$
\item
  3 times 3 sums to update $u_i$ = $9\ FLOPs$.
\end{itemize}

The summation of all four contributions gives a total of 441 operations and by dividing by the memory traffic we
obtain the operational intensity $\mathcal{I}_{stiff}$ for variable stiffness and $\mathcal{I}_{const}$ for constant stiffness:

\begin{equation}
\begin{aligned}
   \mathcal{I}_{stiff} &= \frac{441N}{112N} = 3.93, \\
   \mathcal{I}_{const} &= \frac{441N}{28N} = 15.75.
\end{aligned}
\label{OIrsf}
\end{equation}

Using the OI values derived above we can now quantify the results presented by
\citet{Weiss2013} by interpreting their runtime results with respect to our performance measure. The achieved GFLOPS have been obtained on the basis of 1000 time steps with $8^{th}$ order spatial finite-differences and $2^{nd}$ order
temporal finite-differences. We interpret Fig. 11a) of \citet{Weiss2013} to give
a run time of approximately $53$ seconds and a domain size of $N=225^3$. We obtain
with this parameter the following achieved performances:

\begin{equation}
\begin{aligned}
F &= \frac{N^3 F_{kernel} N_{t}}{W}, \\
  &= \frac{225^3\times441\times1000}{53}, \\
  &= 94.8\text{GFLOPS},
\end{aligned}
\end{equation}
where $N_t$ is the number of time steps, and $W$ is the run time.

\begin{figure}
\centering
\includegraphics[width=1.000\hsize]{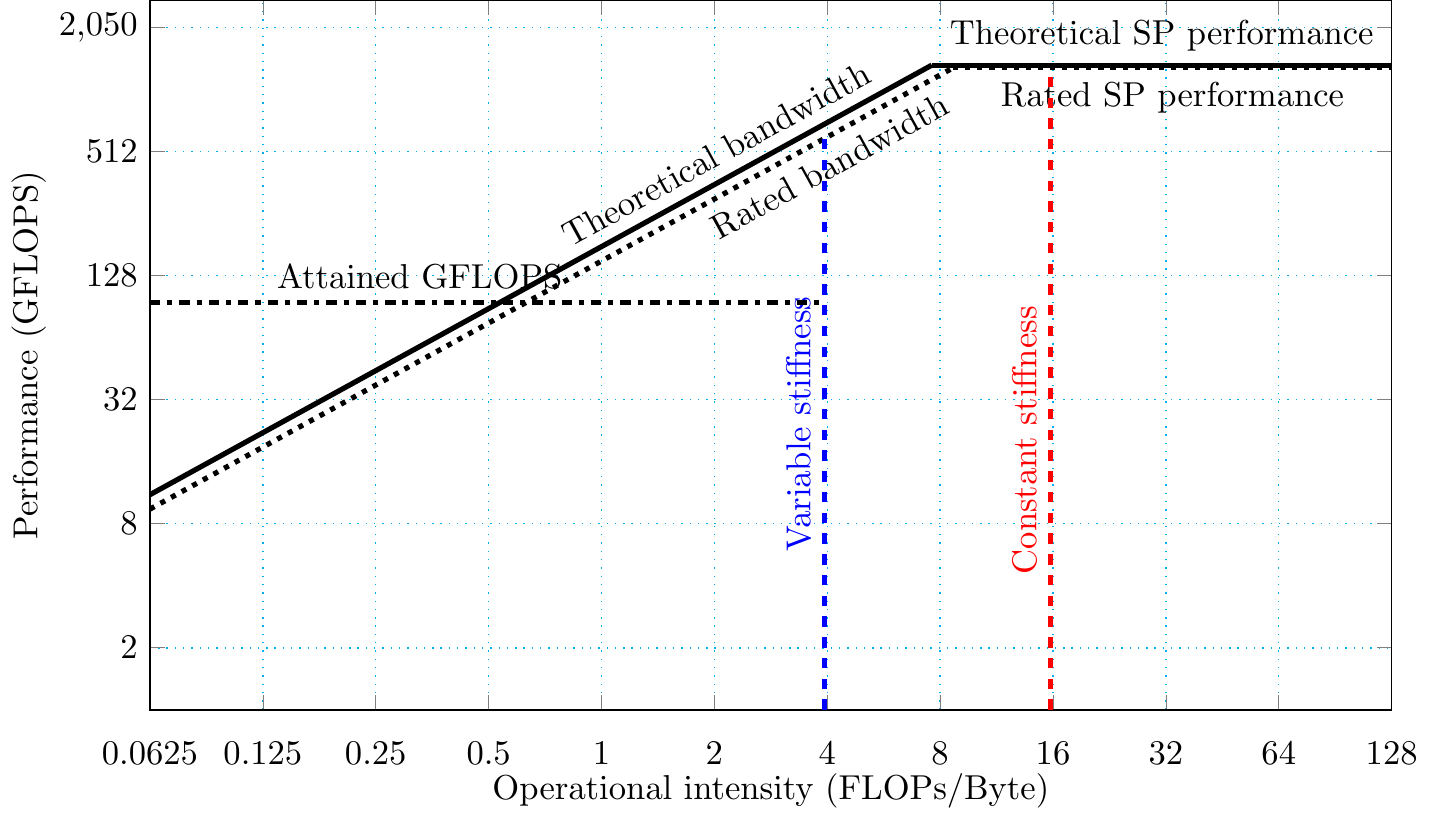}
\caption{Roofline model for the 3D elastic anisotropic kernel from
\citep{Weiss2013} on a 480-core NVIDIA GTX480 GPU (with hardware
specification from \citet{Konstantinidis2015}).}
\label{RSF_roof_GPU}
\end{figure}

Fig.~\ref{RSF_roof_GPU} shows the calculated performance in relation to our
predicted algorithmic bounds $\mathcal{I}_{stiff}$ and $\mathcal{I}_{const}$.
The use of a constant stiffness tensor puts the OI of the considered equation
in the compute-bound region for the benchmarked GPU architecture (NVIDIA
GTX480). Assuming a spatially varying stiffness tensor, we can calculate an achieved hardware
utilization of $40.5\%$ based on the reported results, assuming an achievable
peak memory bandwidth of $150.7\ GByte/s$, as reported by
\citet{Konstantinidis2015} and a maximum achievable performance of $150.7\
GByte/s \times 1.5528\ FLOPs/Byte = 234\ GFLOPS$. Assuming $80\%$ \citep{Andreolli2015} of peak performance is achievable, the roofline model suggests that there is still potential to double the performance of the code through software optimization. It is not possible to draw such a conclusion from traditional performance measures such as timings or scaling plots. This highlights the importance of a reliable performance model that can provide an absolute measure of performance in terms of the algorithm and the computer architecture.

{\color{black} 
\section{Cost-benefit analysis}\label{cost}

So far we discussed the design of finite-difference algorithms purely from a performance point of view without regard to the numerical accuracy and  cost-to-solution. Now we discuss the impact of the discretization order on the achieved accuracy of the solution and how that, in turn, affects the wall clock time required for computation. To do so, we look at the numerical requirements of a time-stepping algorithm for the wave-equation. More specifically we concentrate on two properties, namely dispersion and stability, in the acoustic case. This analysis is extendable to more advanced wave-equations such as VTI and TTI with additional numerical analysis. The dispersion criteria and stability condition for the acoustic wave-equation is given by ~\citep{CFL,dispersion}:

\begin{equation}\label{eq:stability}
	\begin{aligned}
&{v_{max}dt\over h}\le \sqrt{a_1\over a_2} \ \text{CFL condtion, stability} \\
&h\le {v_{min}\over pf_{max}} \ \text{dispersion criterion},
	\end{aligned}
\end{equation}
where:  
\begin{itemize}
\item[${a_1}$] is the sum of the absolute values of the weights of the finite-difference scheme for the second time derivative of the wavefield; ${\partial^2u \over \partial t^2}$
\item[$a_2$] is the sum of the absolute values of the weights of the finite-difference approximation of $\nabla^2u$;
\item[$v_{max}$] is the maximum velocity;
\item[$f_{max}$] is the maximum frequency of the source term that defines the minimum wavelength for a given minimum velocity $\lambda_{min} = \frac{v_{min}}{f_{max}}$;
\item[$p$] is the number of grid points per wavelength. The number of grid points per wavelength impacts the amount of dispersion (different wavelengths propagating at different velocities) generated by the finite-difference scheme. The lower the number, the higher the dispersion will be for a fixed discretization order.
\end{itemize}
These two conditions define the computational setup for a given source and physical model size. 
Knowing that $a_2$ increases with the spatial discretization order,  Eq.~\ref{eq:stability} shows that higher discretization orders require a smaller time-step hence increasing the total number of time steps for a fixed final time and grid size. However, higher order discretizations also allow to use less grid points per wavelength (smaller $p$). A smaller number of grid points per wavelengths leads to a smaller overall computational domain as a fixed physical distance is represented by a coarser mesh and as the grid spacing has been increased, the critical time-step (maximum stable value) is also increased. Overall, high order discretizations have  better computational parameters for a predetermined physical problem.
From these two considerations, we can derive an absolute cost-to-solution estimation for a given model as a function of the discretization order for a fixed maximum frequency and physical model size. The following results are not experimental runtimes but estimations of the minimum achievable runtime assuming a perfect performance implementation.
We use the following setup:

\begin{itemize}
	\item We fix the physical model size as 500 grid point in all three directions for a second order discretization (minimum grid size).
	\item The number of grid points per wavelength is $p=6$ for a second order spatial discretization and $p=2$ for a 24th order discretization and varies linearly for intermediate orders.
	\item The number of time steps is 1000 for the second order spatial discretization and computed according to the grid size/time step for other spatial orders.
\end{itemize}

The hypothetical numerical setup (with $a_1=4$, second order time discretization) is summarized in Table~\ref{tab:stab}. We combine the estimation of a full experimental run with the estimated optimal performance and obtain an estimation of the optimal time-to-solution for a fixed physical problem. The estimated runtime is the ratio of the total number of GFLOPs (multiply $\mathcal{F}_{kernel}$ by the number of grid points and time steps) to the maximum achievable performance for this OI. Table ~\ref{tab:cost} shows the estimated runtime assuming peak performance on two systems: a dual-socket Intel Xeon E5-2697 v2 and an Intel Xeon Phi 7120A co-processor.

\begin{table}[ht] \centering
\begin{tabular}{ccccccc}
Order      & $a_2$ &$p$& $h$ & $dt$ & $N$ & $n_t$ \\
\hline
2nd order  & 12    & 6 & 1   & 0.5774 & 1.25e+08 & 1000 \\
6th order  & 18.13 & 5 & 1.2 & 0.5637 & 7.24e+07 & 1024 \\
12th order & 21.22 & 4 & 1.5 & 0.6513 & 3.70e+07 & 887 \\
18th order & 22.68 & 3 & 2   & 0.8399 & 1.56e+07 & 688 \\
24th order & 23.57 & 2 & 3   & 1.2359 & 4.63e+06 & 468 \\
\end{tabular}
\caption{Cost-to-solution computational setup summary.}
\label{tab:stab}
\end{table}

\begin{table}[ht]
\centering
\begin{adjustbox}{width=1\textwidth}
\begin{tabular}{ccccccc}
Order & $\mathcal{I}_{alg}(k)$ & GFLOPs & GFLOPS Xeon & GFLOPS Phi & Runtime Xeon & Runtime Phi \\
\hline
2nd   & 1.375 & 2.75e+03 & 137.5 & 275 & 20s & 10s\\
6th   & 2.875 & 3.414e+03  & 287.5 & 575 &12s & 6s\\
12th  & 5.125 & 2.691e+03  & 512.5 & 1025 &6s & 3s\\
18th  & 7.375 & 1.266e+03  & 737.5 & 1475 & 2s & 1s\\
24th  & 9.625 & 3.337e+02 & 962.5 & 1925 & 1s& 1s\\
\end{tabular}
\end{adjustbox}
\caption{Cost-to-solution estimation for several spatial discretizations on fixed physical problem.}
\label{tab:cost}
\end{table}

We see that by taking advantage of the roofline results in combination with the stability conditions, we obtain an estimate of the optimal cost-to-solution of an algorithm. It can be seen that higher order stencils lead to better hardware usage by lowering the wall-time-to-solution. These results, however, rely on mathematical results based on homogeneous velocity. In the case of an heterogenous model, high order discretizations may result in inaccurate, even though stable and non dispersive, solutions to the wave-equation. The choice of the discretization order should then be decided with more than just the performance in mind.

}
\section{Conclusions}

Implementing an optimising solver is generally a long and expensive
process. Therefore, it is imperative to have a reliable estimate of the
achievable peak performance, FLOPS, of an algorithm at both the design
and optimised implementation stages of development.

The roofline model provides a readily understandable graphical tool, even for a
non-specialist, to quickly assess and evaluate the computational effectiveness
of a particular implementation of an algorithm.  We have shown how the roofline
model can be applied to finite-difference discretizations of the wave-equation
commonly used in the geophysics community. Although the model is quite simple,
it provides a reliable estimate of the peak performance achievable by a given
finite-difference discretization regardless of the implementation. Not only
does this aid the algorithm designer to decide between different discretization
options but also gives solver developers an absolute measure of the optimality
of a given implementation. The roofline model has also proved extremely useful
in guiding further optimization strategies, since it highlights the limitations
of a particular version of the code, and gives an indication of whether memory
bandwidth optimisations, such as loop blocking techniques, or {\color{black}FLOPs}
optimisations, such as SIMD vectorisation, are likely to improve results. 

However, one should always be mindful of the fact that it does not provide a
complete measure of performance and should be complemented with other metrics,
such as time to solution or strong scaling metrics, to establish a full
understanding of the achieved performance of a particular algorithmic choice
and implementation.


\section{Acknowledgements}

This work was financially supported in part by the Natural Sciences and
Engineering Research Council of Canada Collaborative Research and
Development Grant DNOISE II (CDRP J 375142-08) and the Imperial College
London Intel Parallel Computing Centre. This research was carried out as
part of the SINBAD II project with the support of the member
organizations of the SINBAD Consortium.

\bibliographystyle{elsarticle-num-names}
\bibliography{bib_geophys1}

\appendix
\section{Wave-equations}\label{PDE}

In the following equations $u$ is the pressure field in the case of acoustic
anisotropic while $p,r$ are the split wavefields for the anisotropic case. We
denote by $u(.,0)$ and respectively $p,r$ the value of $u$ for all grid points
at time $t=0$. The physical parameters are $m$ the square slowness, $\epsilon,
\delta$ the Thomsen parameters and $\theta,\phi$ the tilt and azimuth. The main
problem with the TTI case is the presence of transient functions ($cos$, $sin$)
known to be extremely expensive to compute (typically about an order of
magnitude more expensive than an add or multiply). Here we will assume these
functions are precomputed and come from a look-up table, thus only involving
memory traffic In the acoustic anisotropic case the governing equations are:

\begin{equation}
\begin{aligned}
    &m \frac{d^2 u(x,t)}{dt^2} - \nabla^2 u(x,t) =q,  \\
    &u(.,0) = 0, \\
    &\frac{d u(x,t)}{dt}|_{t=0} = 0.
\end{aligned}
\label{eqn:Acou}
\end{equation}

In the anisotropic case we consider the equations describe in
~\citep{liu2009stable}. More advanced formulation have been developed however
this equation allow an explicit formulation on the operational intensity and
simple stencil expression. It is the formulation we are also using in our code
base. In the VTI case the governing equations are:

\begin{equation}
\begin{aligned}
    &m \frac{d^2 p(x,t)}{dt^2} - (1+2\epsilon)D_{xx} p(x,t) - \sqrt{(1+2\delta)} D_{zz} r(x,t) =q,  \\
    &m \frac{d^2 r(x,t)}{dt^2} -  \sqrt{(1+2\delta)}D_{xx}  p(x,t) - D_{zz}  r(x,t) =q,    \\
    &p(.,0) = 0, \\
    &\frac{d p(x,t)}{dt}|_{t=0} = 0, \\
    &r(.,0) = 0, \\
    &\frac{d r(x,t)}{dt}|_{t=0} = 0.
\end{aligned}
\label{eqn:vVTI}
\end{equation}

For TTI the governing equations are:

\begin{equation}
\begin{aligned}  
    &m \frac{d^2 p(x,t)}{dt^2} - (1+2\epsilon)(G_{\bar{x}\bar{x}} +G_{\bar{y}\bar{y}} )p(x,t) - \sqrt{(1+2\delta)}G_{\bar{z}\bar{z}} r(x,t) = q,  \\
    &m \frac{d^2 r(x,t)}{dt^2} -  \sqrt{(1+2\delta)}(G_{\bar{x}\bar{x}} +G_{\bar{y}\bar{y}} ) p(x,t) - G_{\bar{z}\bar{z}} r(x,t) = q,     \\
    &p(.,0) = 0, \\
    &\frac{d p(x,t)}{dt}|_{t=0} = 0, \\
    &r(.,0) = 0, \\
    &\frac{d r(x,t)}{dt}|_{t=0} = 0,
\end{aligned}
\label{eqn:TTI}
\end{equation}

where the rotated differential operators are defined as

\begin{equation}
\begin{aligned}
    G_{\bar{x}\bar{x}}  = & cos(\phi)^2 cos(\theta)^2 \frac{d^2}{dx^2} +sin(\phi)^2 cos(\theta)^2 \frac{d^2}{dy^2}+ \\
                                        & sin(\theta)^2 \frac{d^2}{dz^2} + sin(2\phi) cos(\theta)^2 \frac{d^2}{dx dy} - sin(\phi) sin(2\theta) \frac{d^2}{dy dz} -cos(\phi) sin(2\theta) \frac{d^2}{dx dz} \\
    G_{\bar{y}\bar{y}} = & sin(\phi)^2 \frac{d^2}{dx^2} +cos(\phi)^2  \frac{d^2}{dy^2} - sin(2\phi)^2 \frac{d^2}{dx dy}\\
    G_{\bar{z}\bar{z}} = & cos(\phi)^2 sin(\theta)^2 \frac{d^2}{dx^2} +sin(\phi)^2 sin(\theta)^2 \frac{d^2}{dy^2}+  \\
                                      &cos(\theta)^2 \frac{d^2}{dz^2} + sin(2\phi) sin(\theta)^2 \frac{d^2}{dx dy} + sin(\phi) sin(2\theta) \frac{d^2}{dy dz} +cos(\phi) sin(2\theta) \frac{d^2}{dx dz}. \\ 
\end{aligned}
\label{eqn:DiffOp}
\end{equation}

\end{document}